\documentclass[nofootinbib,aps,preprint]{revtex4}
\pdfoutput=1

\usepackage{amsmath} 
\usepackage{epsfig}
\usepackage{cancel}
\usepackage{color}
\usepackage{amssymb}
\usepackage{pstool}
\usepackage{pstricks}
\usepackage{graphicx}
\usepackage{psfrag}
\usepackage{comment}

\newcommand{\GeV}{~\textrm{GeV}}

\usepackage{xcolor}
\definecolor{red}{rgb}{1.0, 0, 0}
\definecolor{blue}{rgb}{0, 0, 1.0}

\def\beq{\begin{equation}}
\def\eeq{\end{equation}}
\def\beqa{\begin{eqnarray}}
\def\eeqa{\end{eqnarray}}
\def\ben{\begin{enumerate}}
\def\een{\end{enumerate}}

\begin{document}  
  
  
\vspace*{18pt}   
  
\title{\boldmath Cosmological Constraints on MFV SUSY}

\author{Michael Savastio}
\affiliation{Department of Physics, LEPP, Cornell University, Ithaca, NY 14853 \vspace*{8pt}}
  
\begin{abstract} \vspace*{10pt} 
We study cosmological constraints in the context of $R$-parity violating MFV SUSY and find it is driven to $\tan(\beta)\approx1$.  These constraints are from two sources: first from the requirement that baryon number violation not undo baryogenesis and second that the flux of decay products from gravitino dark matter not exceed that observed by experiments such as PAMELA and Fermi LAT.  The latter discussion favors relatively low gravitino masses of a few GeV.
\end{abstract}  
  
\maketitle

\section{Introduction}
As SUSY searches at the LHC shrink the parameter space and endanger naturalness, the need for SUSY models with radically different phenomenology becomes apparent.  One way of achieving this is by relaxing the requirement of $R$-parity \cite{Barbier:2004ez}.  A particularly elegant way of achieving this is with MFV SUSY in which the SM Yukawas are assumed the only source of flavor and $R$-parity violation \cite{Csaki:2011ge,Smith:2008ju,Nikolidakis:2007fc}, enabling a triply Yukawa suppressed $B$ and $R$-parity violating coupling.  This coupling is trilinear in the quark superfields with the generation dependent coefficient
\begin{equation}
\lambda''_{ijk}=w''V^{*}_{il}\epsilon_{jkl}\frac{m^{(u)}_{i}m^{(u)}_{j}m^{(d)}_{k}}{v^{3}}\frac{1}{\sin(\beta)\cos^{2}(\beta)}
\end{equation}  
where $w''$ is, by assumption, an $O(1)$ parameter and $m^{(u)}_{i}$, $m^{(d)}_{i}$ are the up-type and down-type quark masses respectively.  The indices indicate generation number.  Note that for large $\tan(\beta)$ we have $\csc(\beta)\sec^{2}(\beta)\to\tan^{2}(\beta)$.  The form of $\lambda''_{ijk}$ is determined by demanding that the associated operator is a gauge singlet and a singlet under the $SU(3)^{5}$ MFV flavor group (see \cite{D'Ambrosio:2002ex}).  The resulting theory has the advantage of allowing superpartners to decay, thus evading missing energy searches, as well as evading stringent bounds from proton decay, neutron oscillations and flavor physics.  
	
	Since MFV SUSY was formulated specifically with the intention of avoiding collider constraints, it is most often discussed in that context, whether through direct searches or flavor physics.  There is a great deal of literature on the cosmological implications of $R$-parity violation \cite{Baltz:1997ar,Endo:2009cv,Ibarra:2007jz} but none which specifically address the minimal version of MFV SUSY.  (For an extension of MFV SUSY with new, stable DM candidates see \cite{Batell:2013zwa}.)  Since MFV SUSY is itself so constraining (for our purposes there are essentially only two free parameters: $\tan(\beta)$ and $w''$), reviewing the cosmological constraints in this context is enlightening.  
	
	We find that there are two observations which provide significant constraints.  The first is that any coupling which violates baryon number will rapidly bleed off existing baryon number if it is ever in thermal equilibrium, jeopardizing baryogenesis.  Notably, this process is generation independent before the electroweak phase transition.  As we will see this requires the baryon number violating coupling $\lambda''$ to be small as long as baryon number is generated at a temperature above the electroweak phase transition.  A second important observation involves limits on dark matter decay product flux.  In MFV SUSY, since the neutralino is very short lived on cosmic time scales, the most natural DM candidate is the gravitino.  While gravitino production is much the same in MFV SUSY as in other SUSY models, in MFV it will decay via $R$-parity violating couplings.  As we will see, the gravitino will be long lived on cosmic time scales, however the lifetime will nevertheless be short enough to produce a significant abundance of anti-protons and $\gamma$-rays which have not yet been observed.  This would seem to imply that if MFV SUSY is a realistic model of nature and if gravitinos are to indeed play the part of DM, we should be on the verge of detecting them indirectly.

\section{Baryogenesis Constraint on $\lambda''$}
Any quantum number which is odd under $CPT$ will be rapidly driven to zero in thermal equilibrium if it is not strictly conserved.  It is therefore possible that the $R$-parity and baryon number violating coupling $\lambda''$ can destroy existing baryon number, potentially undoing baryogenesis.  Any time the rate of the interaction mediated by $\lambda''$ is faster than the rate of the expansion of the universe it will be in equilibrium.  Since by dimensional analysis this rate is $\Gamma\sim|\lambda''|^{2}T$ at temperatures above the electroweak scale, and the rate of expansion is $H\sim T^{2}/M_{P}$, the rate of expansion decreases faster than the rate of interaction, so the baryon number violating interaction will be in equilibrium at late times.  At the electroweak phase transition the quarks participating in the $B$-violating process gain masses, with the top and soon the bottom masses of the same order as the temperature so that the most dominant $B$-violating interactions, those with the top, fall out of equilibrium.  Therefore, the rate of the $B$-violating process will fall off rapidly after the electroweak phase transition.  It is therefore sufficient to demand that $\lambda''$ is not in equilibrium by $T\sim100\:\textrm{GeV}$.  In \cite{Endo:2009cv} a Boltzmann equation evolution of the baryon density was used to derive the constraint
\begin{equation}
\sqrt{\sum_{ijk}|\lambda''_{ijk}|^{2}}\lesssim 4\cdot10^{-7}
\end{equation}
for squarks $m_{\tilde{q}}=200\GeV$.  Interestingly, this bound is nearly independent of the squark masses, for $m_{\tilde{q}}=1200\GeV$ the bound rises to only $5\cdot10^{-7}$, so this does not affect our estimate.  Details including explicit Boltzmann equations can be found in \cite{Endo:2009cv}.  
Now we see that there is a stringent baryogenesis constraint on the product $w''\tan^{2}(\beta)$ at large $\tan(\beta)$.  Note that
\begin{equation}
\sqrt{\sum_{ijk}|\lambda''_{ijk}|^{2}}\sim(2\cdot10^{-7})\frac{w''}{\sin(\beta)\cos^{2}(\beta)}
\end{equation}
This implies, for $\tan(\beta)$ not below or too close to $1$
\begin{equation}
\tan(\beta)\lesssim\sqrt{\frac{2}{w''}}
\end{equation}
For $w''=10^{-1}$ this implies $\tan(\beta)\lesssim4$.  This is quite significant since the requirement that the top Yukawa coupling $y_{t}$ not become non-perturbatively large above the electroweak scale gives $\tan(\beta)\gtrsim1.2$ \cite{oai:arXiv.org:hep-ph/9709356}.  $w''=1$ is therefore very nearly ruled out.

	This constraint is quite important since MFV models are usually considered with large $\tan(\beta)$ to provide more realistic collider scenarios.  Indeed, it puts tension on some of the most natural LSP candidates since they would then be extremely long lived.  Following the estimates in \cite{Csaki:2011ge} we have for the stop
\begin{equation}
\tau_{\tilde{t}}\sim (2\:\textrm{cm})\frac{1}{w''^{2}}\left(\frac{1}{\tan(\beta)}\right)^{4}\left(\frac{300\GeV}{\tilde{m}_{t}}\right)
\label{stop}
\end{equation}
for the left-handed sbottom
\begin{equation}
\tau_{\tilde{b}_{L}}\sim (41\:\textrm{m})\frac{1}{w''^{2}}\left(\frac{1}{\tan(\beta)}\right)^{6}\left(\frac{300\GeV}{\tilde{m}_{b_{L}}}\right)
\label{sbottom}
\end{equation}
and for the neutralino
\begin{equation}
\tau_{\tilde{N}}\sim (2\:\textrm{m})\frac{1}{w''^{2}}\left(\frac{1}{\tan(\beta)}\right)^{4}\left(\frac{300\GeV}{m_{\tilde{N}}}\right)
\label{neutralino}
\end{equation}
As we see, the sbottom is ruled out as an NLSP for small $\tan(\beta)$ (the exclusion comes from heavy stable charged particle (HSCP) searches \cite{Khachatryan:2011ts,Aad:2011hz})  while the neutralino is marginal.  Requiring that the stop decays within $5~\textrm{m}$ so that it is too short lived to travel all the way through the muon arms of CMS and ATLAS and has a chance of evading HSCP searches we have
\begin{equation}
\tan(\beta)\gtrsim\frac{1}{4}\sqrt{\frac{1}{w''}}\left(\frac{300\GeV}{\tilde{m}_{t}}\right)^{1/4}
\label{collider_constraint}
\end{equation}  
Even for $w''=10^{-1}$ (which gives $\tan(\beta)\gtrsim 3/4$) this bound is weaker than that from the perturbativity requirement on $y_{t}$, but as we will see it may be significant for very low $w''$ which baryogenesis and DM constraints force us to consider.  For example, for $w''=10^{-2}$ we have $\tan(\beta)\gtrsim2.5$.  Recall also that the left hand side of (\ref{collider_constraint}) should be replaced with $\sqrt{\csc(\beta)\sec^{2}(\beta)}$ for $\tan(\beta)\lesssim1$.  Later (see Fig. \ref{exclusion0}) we will review this in relation to cosmological constraints.

	Of course, the baryogenesis constraint can be avoided either in part or entirely in any scenario in which baryon number is generated at or below the electroweak phase transition.  There would likely be some constraint on $\lambda''$ in electroweak barygoenesis since baryon number is still generated in the symmetric phase, but we expect it to be much weaker.  (Electroweak baryogenesis may have its own implications on $\tan(\beta)$, see \cite{Cline:1996cr}.)  Scenarios with arbitrarily low reheat temperature such as Affleck-Dine baryogenesis (for a review see \cite{Dine:2003ax}) evade this constraint entirely.  If the recent BICEP2 results \cite{Ade:2014xna} are confirmed inflation happens at GUT scales, making it much more difficult to achieve very low reheat temperatures so that our baryogenesis constraint is harder to avoid using Affleck-Dine.  In other scenearios in which baryon number is generated above the electroweak phase transition such as thermal leptogenesis, this constraint remains quite severe.

\section{Gravitino DM in MFV SUSY}

\begin{figure}[t]
\begin{center}
\includegraphics[height=80mm,width=120mm]{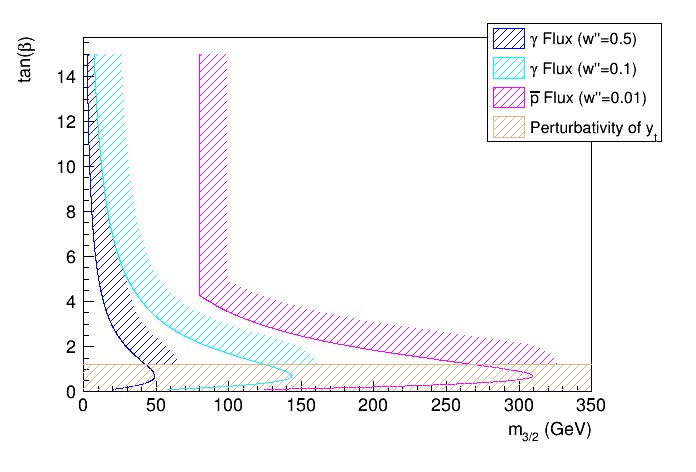}
\caption{Upper limit on the gravitino mass as a function of $\tan(\beta)$ in order for it not to produce excessive $\bar{p}$ or $\gamma$ flux.  Excluded regions are on the sides of the solid lines with hashing.  The reason for the hard cutoff at $m_{3/2}=m_{W}$ in the $w''=0.01$ line is because the $\bar{p}$ flux constraint is based on $\tilde{G}\to W^{\pm}\ell^{\mp}$.\label{exclusion_gravmass}}
\end{center}
\end{figure}

In MFV SUSY gravitinos are unstable, and if they are the LSP and lighter than the top they decay predominantly through the channel $\tilde{G}\to cbs$ with lifetime
\begin{equation}
\tau_{3/2}\sim(2\cdot10^{25}~\textrm{s})\frac{1}{w''^{2}}\left(\frac{100\GeV}{m_{3/2}}\right)^{3}\sin^{2}(\beta)\cos^{4}(\beta)
\end{equation}
(For a review on gravitino coupling to matter, see, for example \cite{Moroi:1995fs}.)  Again, for large $\tan(\beta)$, $\sin^{2}(\beta)\cos^{4}(\beta)\to \tan^{-4}(\beta)$.  Note that if the gravitino is heavier than the top, it will not be sufficiently long-lived to be a DM candidate in MFV SUSY.  While our estimate of $\tau_{3/2}$ is indeed much longer than the age of the universe, it will lead to a significant excess of cosmic rays.  Note that since $cbs$ has baryon number $|B|=1$, the ultimate final state will necessarily contain at least one proton or anti-proton.  There will also be a significant number of photons produced in the Dalitz decays of neutral pions.  

\subsection{Anti-proton Constraints from PAMELA}
\begin{table}
\begin{tabular}{l|c|c|c}
 & ~~$cbs$~~ & ~~$Z\nu$~~ & ~~$W^{\pm}\ell^{\mp}$~~ \\
\hline
$p+\bar{p}$ multiplicity & $1.9$ & $1.5$ ($1.7$) & $1.5$ ($1.6$) 
\end{tabular}
\caption{Multiplicities of protons or anti-protons for the final state relevant to MFV ($cbs$) and two others, from Pythia 8, for $100\GeV$ gravitinos.  In parentheses are multiplicities reported by \cite{Delahaye:2013yqa} to which we compare.  We attribute the difference to the different versions of Pythia being used.  We find that these values are nearly independent of the gravitino mass in the range of interest.}
\label{proton_multiplicity}
\end{table}

\begin{figure}[t]
\begin{center}
\includegraphics[height=60mm,width=80mm]{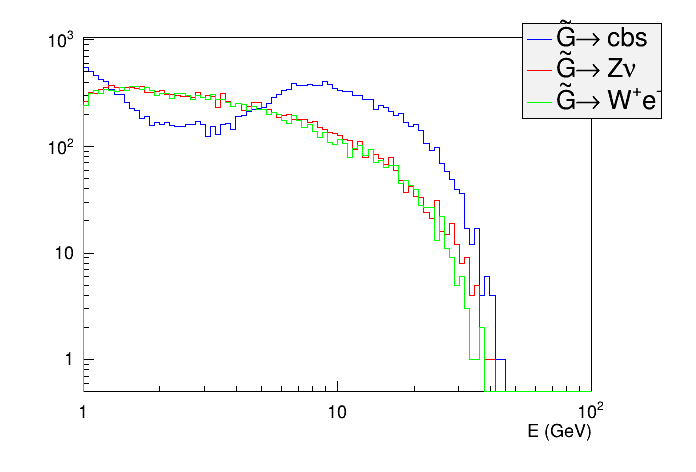}
\includegraphics[height=60mm,width=80mm]{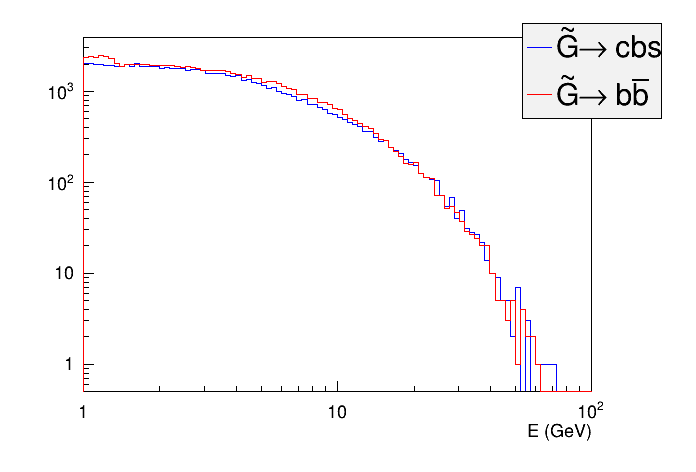}
\caption{Comparison of anti-proton (left) and photon (right) spectra for various different final states generated using Pythia 8 and $m_{3/2}=100\GeV$.  The ordinate shows the number of particles, where $10^{4}$ events were generated for each case.  The shapes of the spectra were found to depend weakly on the gravitino mass in the region of interest (down to about $m_{W}$ for anti-protons and about $20\GeV$ for photons). \label{spectra}}
\end{center}
\end{figure}

	The PAMELA experiment reports no statistically significant excess of anti-protons in the range $60~\textrm{MeV}$ to $180~\textrm{GeV}$ after 850 days of running \cite{Adriani:2010rc}.  A number of analyses have been carried out to derive lower limits on gravitino DM lifetime based on this data \cite{Cholis:2010xb,Garny:2012vt,Delahaye:2013yqa,Cirelli:2013hv}, some of which study RPV scenarios, though not for the decay mode $\tilde{G}\to cbs$.  Unfortunately these bounds suffer from enormous uncertainties of several orders of magnitude (in seconds) due to astrophysical effects on the propagation of anti-protons through the galaxy as well as astrophysical background and uncertainty in the shape of the galactic DM halo profile.  For extreme values of astrophysical parameters, all these analyses conclude a lower limit of no less than $10^{26}~\textrm{s}$ for the DM lifetime with masses in the range $m_{W}$ to about $m_{t}$ (unfortunately these do not consider very low, order $\GeV$ gravitino masses due to the final states being considered, though we will see that the resulting bounds force us into this region).  Though they consider different decay modes, the $cbs$ final state has a similar $p+\bar{p}$ multiplicity to those studied (see Table \ref{proton_multiplicity}).  For example \cite{Delahaye:2013yqa} considers $W\ell$ and $Z\nu$ final states with $p+\bar{p}$ multiplicities each of about $1.6$, and concludes a lower limit on the lifetime of about $2\cdot10^{27}~\textrm{s}$ for $100\GeV$ gravitinos (and roughly similar values up to the top mass).  We used simulations in Pythia 8 \cite{Sjostrand:2007gs} to conclude that in our case the anti-proton multiplicities and spectra are similar to the analysis of \cite{Delahaye:2013yqa} (see Fig. \ref{spectra}).  One should note that the PAMELA data is most constraining in the range from a few $\GeV$ through a few times $10\GeV$.  There is a dip in the $\tilde{G}\to cbs$ anti-proton spectrum around $2\GeV$, but it is greater than the $\tilde{G}\to Z\nu (W^{\pm}e^{\mp})$ spectra above about $10\GeV$.   We can therefore repurpose the $\tilde{G}\to Z\nu$ and $\tilde{G}\to W^{\pm}e^{\mp}$ analyses to conclude that it would place a lower limit on the $\tilde{G}\to cbs$ lifetime of about 
\begin{equation}
\tau_{3/2}\gtrsim 10^{27}~\textrm{s},
\label{anti-proton}
\end{equation}
restricting us to a rather uncomfortable region of MFV SUSY parameter space.  While one can increase the lifetime to about $10^{27}~\textrm{s}$ by taking $w''\sim0.1$ for $\tan(\beta)\approx1$, in doing so one starts to create tension with collider data since the NLSP's may become so long-lived as to be excluded by missing energy or direct searches as can be seen in Eqs. (\ref{stop},\ref{sbottom},\ref{neutralino}).  This lifetime limit gives us a constraint on $w''\tan^{2}(\beta)$ for $\tan(\beta)\gtrsim1$
\begin{equation}
\tan(\beta)\lesssim\frac{0.4}{\sqrt{w''}}\left(\frac{100\GeV}{m_{3/2}}\right)^{3/4}
\end{equation}
In Fig. \ref{exclusion_gravmass} we plot upper limits on $m_{3/2}$ as a function of $\tan(\beta)$ derived from (\ref{anti-proton}).
	
\subsection{$\gamma$-Ray Constraints from Fermi LAT}
	Simpler analyses are carried out searching for an excess of high energy cosmic photons due to dark matter decay or annihilation, most recently from Fermi LAT \cite{Grefe:2011dp,Huang:2011xr,Mazziotta:2012ux}.  Even in these cases it is not quite so simple to compare with the predicted flux because of the details of the data selection (a glance at our references will reveal that this can still be the cause of significant uncertainty), however we can still take advantage of lifetime limits concluded from these analyses.   
	
\begin{table}	
\begin{tabular}{c|c|c}
 & ~~$cbs$~~ & ~~$b\bar{b}$~~\\
\hline
$\gamma$ multiplicity & $15.7$ & $16.7$
\end{tabular}
\caption{Comparison of multiplicities of final state photons from $cbs$ and from $b\bar{b}$ as determined from Pythia 8 for $100\GeV$ gravitinos.  Again, we find these values to be approximately independent of $m_{3/2}$ in the region of interest (for $m_{3/2}=20\GeV$ we have the $cbs$ and $b\bar{b}$ multiplicities at about $8$ and $13$ respectively).}
\label{photon_multiplicity}
\end{table}

	We compare photons produced in $\tilde{G}\to cbs$ to those produced in $\tilde{G}\to b\bar{b}$ as in \cite{Huang:2011xr}.  Using Pythia 8 we find that the photon multiplicities of these final states are similar (see Table \ref{photon_multiplicity}).  We also find that the spectra of photons in these cases are very nearly identical (see Fig. \ref{spectra}).  From this and \cite{Huang:2011xr} we can conclude the somewhat weaker bound
\begin{equation}
\tau_{3/2}\gtrsim 10^{26}~\textrm{s}
\end{equation}

\subsection{PAMELA $e^{+}$ Excess}
\begin{table}	
\begin{tabular}{c|c|c|c}
 & ~~$cbs$~~ & ~~$Z\nu$~~ & ~~$W^{\mp}e^{\pm}$~~\\
\hline
$e^{+}+e^{-}$ multiplicity & $0.48$ & $0.44$ & $1.39$
\label{posMultiplicity}
\end{tabular}
\caption{Comparison of multiplicities of final state electrons and positrons from $cbs$, $Z\nu$ and $W^{\pm}e^{\mp}$, found generating $10^{4}$ events in Pythia 8 with $m_{3/2}=100\GeV$.}  
\end{table}
We should briefly mention that MFV SUSY gravitinos cannot explain the PAMELA and AMS positron excess \cite{Aguilar:2013qda,Adriani:2013uda}.  An analysis has been carried out showing that gravitino decay to $Z\nu$ or $W^{\pm}e^{\mp}$ can explain this excess \cite{Ibarra:2008jk}.  While these final states have similar overall positron multiplicities to $cbs$ (see Table \ref{posMultiplicity}), both $Z\nu$ and $W^{\pm}e^{\mp}$ exhibit a sharp rise in their spectra around a few tens of GeV (for $100\GeV\leq m_{3/2}\leq200\GeV$).  More importantly it can be seen in \cite{Ibarra:2008jk} that gravitino masses at least as large as the top mass are need to explain the entire excess, while also requiring $\tau_{3/2}\sim10^{26}\:\textrm{s}$, which is essentially ruled out in this model by the absence of an anti-proton excess.

\section{conclusion}

\begin{figure}[t]
\begin{center}
\includegraphics[height=80mm,width=120mm]{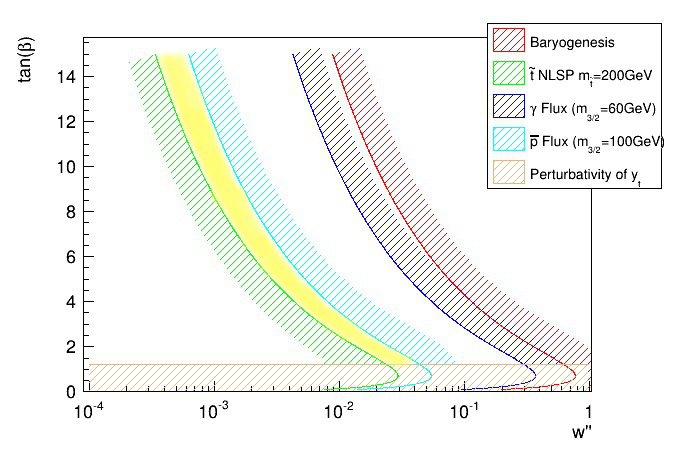}
\includegraphics[height=80mm,width=120mm]{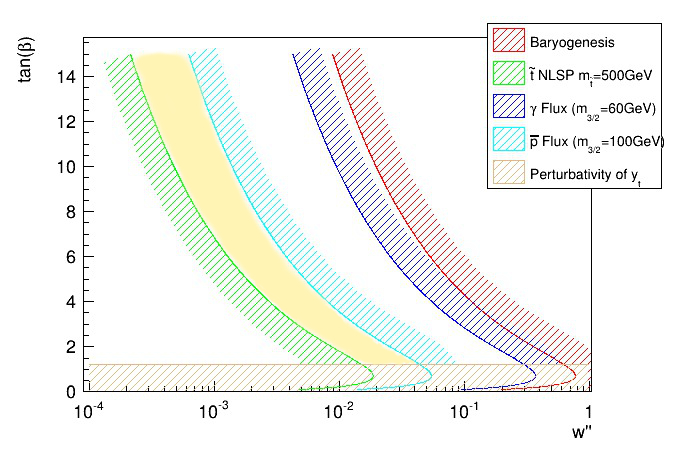}
\caption{Constraints on the MFV SUSY parameter space due to cosmological considerations.  Excluded regions are on the sides of the lines with hashing.  The region allowed by all constraints is highlighted in yellow.  We show a possible bound from below assuming the $\tilde{t}$ is the NLSP and requiring it not live more than $5~\textrm{m}$ so that it would not have been detected as a heavy stable charged particle at the LHC for two different values of $m_{\tilde{t}}$.  (The limit of $5~\textrm{m}$ is taken so large in order to account for the possibility of low-velocity stops and is thought to be quite conservative.) \label{exclusion0}}
\end{center}
\end{figure}

We have seen that cosmological evidence combined with the requirement that the NLSP not be too long lived severely constrains MFV SUSY, at least in the most orthodox models of reheating and baryogenesis.  These constraints are shown on the $(w'',\tan(\beta))$ parameter space in Fig. \ref{exclusion0}.  This combined with flavor physics constraints \cite{Arbey:2012ax,Nir:2007xn} would seem to strongly suggest that $\tan(\beta)$ must be small even for MFV SUSY, and it is pushed uncomfortably close to the limit by requiring $y_{t}$ remains perturbative $\tan(\beta)\gtrsim1.2$.  Certainly, if the gravitino DM scenario is to be believed, then combined with baryogenesis considerations it is certainly true that $\lambda''$ must be significantly smaller than its upper limit based on proton decay and flavor physics alone.  The gravitino is still a DM candidate, since there is no obstacle in making $m_{3/2}$ as low as $10\GeV$ in order to avoid constraints from indirect detection, but to do so one must be willing to accept a reheating temperature of no more than about $10^{7}\GeV$, which would be problematic for thermal leptogenesis.
The cosmological constraints we have considered here suggest a lower gravitino mass than is often recently studied.

\acknowledgments{
The author thanks Maxim Perelstein and Yuval Grossman for useful discussion and advice.  The work of MS is supported in part by the U.S. National
Science Foundation.}



\end{document}